# Bayes, Jeffreys, Prior Distributions and the Philosophy of Statistics[1]

**Andrew Gelman**

I actually own a copy of Harold Jeffreys's *Theory of Probability* but have only read small bits of it, most recently over a decade ago to confirm that, indeed, Jeffreys was not too proud to use a classical chi-squared *p*-value when he wanted to check the misfit of a model to data (Gelman, Meng and Stern, 2006). I do, however, feel that it is important to understand where our probability models come from, and I welcome the opportunity to use the present article by Robert, Chopin and Rousseau as a platform for further discussion of foundational issues.[2]

In this brief discussion I will argue the following: (1) in thinking about prior distributions, we should go beyond Jeffreys's principles and move toward weakly informative priors; (2) it is natural for those of us who work in social and computational sciences to favor complex models, contra Jeffreys's preference for simplicity; and (3) a key generalization of Jeffreys's ideas is to explicitly include model checking in the process of data analysis.

## THE ROLE OF THE PRIOR DISTRIBUTION IN BAYESIAN DATA ANALYSIS

At least in the field of statistics, Jeffreys is best known for his eponymous prior distribution and, more generally, for the principle of constructing noninformative, or minimally informative, or objective, or reference prior distributions from the likelihood (see, for example, Kass and Wasserman, 1996). But it can notoriously difficult to choose among noninformative priors; and, even more importantly, seemingly noninformative distributions can sometimes have strong and undesirable implications, as I have found in my own experience (Gelman, 1996, 2006). As a result I have become a convert to the cause of *weakly informative priors*, which attempt to let the data speak while being strong enough to exclude various "unphysical" possibilities which, if not blocked, can take over a posterior distribution in settings with sparse data—a situation which is increasingly present as we continue to develop the techniques of working with complex hierarchical and nonparametric models.

## HOW THE SOCIAL AND COMPUTATIONAL SCIENCES DIFFER FROM PHYSICS

Robert, Chopin and Rousseau trace the application of Ockham's razor (the preference for simpler models) from Jeffreys's discussion of the law of gravity through later work of a mathematical statistician (Jim Berger), an astronomer (Bill Jefferys) and a physicist (David MacKay). From their perspective, Ockham's razor seems unquestionably reasonable, with the only point of debate being the extent to which Bayesian inference automatically encompasses it.

My own perspective as a social scientist is completely different. I've just about never heard someone in social science object to the *inclusion* of a variable or an interaction in a model; rather, the

*Andrew Gelman is Professor, Department of Statistics and Department of Political Science, Columbia University e-mail: gelman@stat.columbia.edu; URL: http://www.stat.columbia.edu/~gelman.*

[1]Discussion of "Harold Jeffreys's Theory of Probability revisited," by Christian Robert, Nicolas Chopin, and Judith Rousseau, for *Statistical Science*.

[2]On the topic of other books on the foundations of Bayesian statistics, I confess to having found Savage (1954) to be nearly unreadable, a book too much of a product of its time in its enthusiasm for game theory as a solution to all problems, an attitude which I find charming in the classic work of Luce and Raiffa (1957) but more of annoyance in a book of statistical methods. When it comes to Cold War-era foundational work on Bayesian statistics, I much prefer the work of Lindley, in his 1965 book and elsewhere.

Also, I would be disloyal to my coauthors if I did not report that, despite what is said in the second footnote in the article under discussion, there is at least one other foundational Bayesian text of 1990s vintage that continues to receive more citations than Jeffreys.

most serious criticisms of a model involve worries that certain potentially important factors have *not* been included. In the social science problems I've seen, Ockham's razor is at best an irrelevance and at worse can lead to acceptance of models that are missing key features that the data could actually provide information on. As such, I am no fan of methods such as BIC that attempt to justify the use of simple models that do not fit observed data. Don't get me wrong—*all the time* I use simple models that don't fit the data—but no amount of BIC will make me feel good about it![3]

I much prefer Radford Neal's line from his Ph.D. thesis:

> Sometimes a simple model will outperform a more complex model... Nevertheless, I [Neal] believe that deliberately limiting the complexity of the model is not fruitful when the problem is evidently complex. Instead, if a simple model is found that outperforms some particular complex model, the appropriate response is to define a different complex model that captures whatever aspect of the problem led to the simple model performing well.

This is not really a Bayesian or a non-Bayesian issue: complicated models with virtually unlimited nonlinearity and interactions are being developed using Bayesian principles. See, for example, Dunson (2006) and Chipman, George and McCulloch (2008). To put it another way, you can be a practicing Bayesian and prefer simpler models, or be a practicing Bayesian and prefer complicated models. Or you can follow similar inclinations toward simplicity or complexity from various non-Bayesian perspectives.

My point here is only that the Ockhamite tendencies of Jeffreys and his followers up to and including MacKay may derive, to some extent, from the simplicity of the best models of physics, the sense that good science moves from the particular to the general—an attitude that does *not* fit in so well with modern social and computational science.

## BAYESIAN INFERENCE VS. BAYESIAN DATA ANALYSIS

One of my own epiphanies—actually stimulated by the writings of E. T. Jaynes, yet another Bayesian physicist—and incorporated into the title of my own book on Bayesian statistics, is that sometimes the most important thing to come out of an inference is the rejection of the model on which it is based. Data analysis includes model building and criticism, not merely inference. Only through careful model building is such definitive rejection possible. This idea—the comparison of predictive inferences to data—was forcefully put into Bayesian terms nearly thirty years ago by Box (1980) and Rubin (1984) but is even now still only gradually becoming standard in Bayesian practice.

A famous empiricist once said, "With great power comes great responsibility." In Bayesian terms, the stronger we make our model—following the excellent precepts of Jeffreys and Jaynes—the more able we will be to find the model's flaws and thus perform scientific learning.

To roughly translate into philosophy-of-science jargon: Bayesian inference within a model is "normal science," and "scientific revolution" is the process of checking a model, seeing its mismatches with reality, and coming up with a replacement. The revolution is the glamour boy in this scenario, but, as Kuhn (1962) emphasized, it is only the careful work of normal science that makes the revolution possible: the better we can work out the implications of a theory, the more effectively we can find its flaws and thus learn about nature.[4] In this chicken-and-egg process, both normal science (Bayesian inference) and revolution (Bayesian model revision) are useful, and they feed upon each other. It is in this sense that graphical methods and exploratory data analysis can be viewed as explicitly Bayesian, as tools for comparing posterior predictions to data (Gelman, 2003).

To get back to the Robert, Chopin, and Rousseau article: I am suggesting that their identification (and Jeffreys's) of Bayesian data analysis with Bayesian *inference* is limiting and, in practice, puts an unrealistic burden on any model.

---

[3]See Gelman and Rubin (1995) for a fuller expression of this position, and Raftery (1995) for a defense of BIC in general and in the context of two applications in sociology.

[4]As Kuhn may very well have written had he lived long enough, scientific progress is fractal, with episodes of normal science and mini-revolutions happening over the period of minutes, hours, days, and years, as well as the more familiar examples of paradigms lasting over decades or centuries.

## CONCLUSION

If you wanted to do foundational research in statistics in the mid-twentieth century, you had to be bit of a mathematician, whether you wanted to or not. As Robert, Chopin, and Rousseau's own work reveals, if you want to do statistical research at the turn of the twenty-first century, you have to be a computer programmer.

The present discussion is fascinating in the way it reveals how many of our currently unresolved issues in Bayesian statistics were considered with sophistication by Jeffreys. It is certainly no criticism of his pioneering work that it has been a springboard for decades of development, most notably (in my opinion) involving the routine use of hierarchical models of potentially unlimited complexity, and with the recognition that much can be learned by both the successes and the failures of a statistical model's attempt to capture reality. The Bayesian ideas of Jeffreys, de Finetti, Lindley, and others have been central to the shift in focus away from simply modeling data collection and toward the modeling of underlying processes of interest—"prior distributions," one might say.

## ACKNOWLEDGMENTS

We thank Hal Stern for helpful comments and the National Science Foundation for partial support of this research.